# Insulator-metal transition in liquid hydrogen and deuterium

Shuqing Jiang[1,2], Nicholas Holtgrewe[2,3,†], Zachary M. Geballe[2], Sergey S. Lobanov[2,4],

Mohammad F. Mahmood[3], R. Stewart McWilliams[5], Alexander F. Goncharov[1,2]*

[1]Key Laboratory of Materials Physics, Institute of Solid State Physics, Chinese Academy of Sciences, Hefei, Anhui 230031, China.

[2]Geophysical Laboratory, Carnegie Institution of Washington, Washington, DC 20015, USA.

[3]Department of Mathematics, Howard University, 2400 Sixth Street NW, Washington D.C. 20059, USA.

[4] GFZ German Research Center for Geosciences, Section 4.3, Telegrafenberg, 14473 Potsdam, Germany

[5] School of Physics and Astronomy and Centre for Science at Extreme Conditions, University of Edinburgh, Edinburgh EH9 3FD, UK

†Current address: Center for Advanced Radiation Sources, University of Chicago, Chicago, Illinois 60637, USA.

* Corresponding author: agoncharov@carnegiescience.edu

**The insulator-to-metal transition in dense fluid hydrogen is an essential phenomenon to understand gas giant planetary interiors and the physical and chemical behavior of highly compressed condensed matter. Using fast laser spectroscopy techniques to probe hydrogen and deuterium precompressed in a diamond anvil cell and laser heated on microsecond timescales, we observe an onset of metal-like reflectivity in the visible spectral range at P>150 GPa and T≥3000 K. The reflectance increases rapidly with decreasing photon energy indicating free-electron metallic behavior with a plasma edge in the visible spectral range at high temperatures. The reflectivity spectra also suggest much longer electronic collision time (≥1 fs) than previously inferred, implying that metallic hydrogens at the conditions studied are not in the regime of saturated conductivity (Mott-Ioffe-Regel limit). Combined with previously reported data, our results suggest the existence of a semiconducting intermediate fluid hydrogen state *en route* to metallization.**

The insulator-to-metal transition (IMT) in hydrogen is one the most fundamental problems in condensed matter physics [1]. In spite of seeming simplicity of hydrogen (2p+2e in the molecule), the behavior of this system at high compression remains poorly understood. The structural, chemical, and electronic properties of hydrogen and other molecular system are strongly dependent on pressure (density); at high pressures the stability of atomic configurations increases due to an increase in the kinetic energy thus easing the transformation to a metallic state [2,3]. The greatest challenge is to understand the intermediate paired, mixed, and monatomic states, both solid and fluid[3,4]. Of particular interest are the mechanism of IMT, the pressure-temperature (*P-T*) conditions, the existence and location of critical and triple points related to a change in the transition character, and implications to high-temperature superconductivity and the internal structure, composition, temperature, and magnetic fields of gas giant planets[5-7]. Currently, the IMT in hydrogen is expected to occur in two regimes: at low T (<600 K) in the dense solid, where quantum effects are expected to dominate and at high temperatures in the fluid state, where classical entropy must play an important role. In the former scenario there is a possibility of quantum melting where solid $H_2$ melts into a metallic quantum fluid [8,9]. The nature of the metallic fluid at high temperatures, as relevant to planetary interiors, also remains to be established, with questions persisting about electronic transport properties such as electrical conductivity [10-16] and the related chemical state [16,17].

The IMT in fluid hydrogen was initially predicted as the first-order transition ending in a critical point at very high temperatures (10-17 kK) [5-7]. However, dynamic gas gun and laser driven experiments probing changes in electrical conductivity and optical reflectivity found a continuous transition to a metallic state at 50-140 GPa [10,18,19] at lower temperatures, implying a critical temperature below 3 kK. First principles theoretical calculations suggest values of ~2 kK but yield very different critical pressures, and correspondingly positions of the transformation line[20-24]. Arguably, the coupled electron-ion Monte Carlo calculations[17] provide the most accurate predictions; they suggest that the dissociation and metallization transitions coincide (c.f. Ref. [16]), and the critical point is located near 80-170 GPa and 1600-3000 K. While dynamic compression experiments, which explore a variety of *P-T* pathways, agree on existence of the metallic states detected via electrical or optical measurements [10,12,13,18,19], lower temperature data show inconsistent results on the position of metallization and the reflectance of intermediate states[12,13].

Static diamond anvil cell (DAC) experiments combined with laser heating probing similar low temperature fluid states have also yielded controversial results on the electronic properties of hydrogen and the location of the phase lines[11,25,26,27-29]. The difficulty of interpreting these optical DAC experiments is due to indirect probing of the state of hydrogen[27,28], or detection of reflectivity signals superimposed with those of other materials in the DAC cavity[11,26]. The latter results, which reported transient reflectivity and transmission at a few laser wavelengths, have been found inconsistent with the proposed IMT observation[30]. One of the major drawbacks of the majority of preceding dynamic and static experiments is an extreme paucity of robust spectroscopic observations, which are critical for assessing the material electronic properties.

In this paper, we address the challenges raised above by exploring experimentally the electronic states of hydrogen and deuterium in the *P-T* range where the IMT was previously reported but not sufficiently characterized. To overcome the challenges in sustaining hydrogen at these conditions and probing it spectroscopically we applied microsecond single- to several- pulse laser heating in combination with pulsed broadband laser probing. We show that the transition in *P-T* space includes several stages where hydrogen transforms from a transparent insulating state, to an optically absorptive narrow-gap semiconducting state, and finally to a metallic state of high reflectivity. The measured temperatures for the appearance of the metallic fluid are broadly consistent with results of shock experiments at similar pressure[10] but are higher than the conditions for metallization concluded from prior static experiments[11,26,27]. The metallic state exhibits a plasma edge in the visible spectral range, implying a plasma frequency and electronic scattering time that contrasts with previous inferences[12,18,19], mainly based on the Mott-Ioffe-Regal (MIR) limit approximation in which the electronic mean-free-path reaches the interatomic spacing, and in stark disagreement with the prior static experiments probing hydrogen at few laser wavelengths[11].

As in our previously reported experiments in $H_2$[25] and $N_2$[31], here we combine single pulse laser heating of $H_2$ and $D_2$ compressed in the DAC with time-resolved optical emission, transmission and reflectance spectroscopy in the visible spectral range (480-800 nm) using a streak-camera. These microseconds (µs) long experiments (Supplementary Fig. 1) are of durations that are typically sufficient for reaching thermodynamic equilibrium in fluid samples [32]; they are comparable to or longer than dynamic compression experiments commonly considered to reach thermodynamic equilibrium[10]. The *P-T* paths along which we probed the material properties (by varying the heating laser energy at a given nominal pressure) can be considered nearly isobaric, with the maximum added thermal pressure less than 5 GPa at 3000 K[33]. Initial pressure in the hydrogen isotope samples was 124-172 GPa.

A strong extinction of the transmitted light was detected when hydrogen was laser heated above a certain threshold laser heating power (Supplementary Fig. 2). The transient absorption reaches a maximum shortly after the arrival of the heating pulse, followed by a regaining of the transmitted signal. The absorption spectra, measurable only at lower temperatures where transmission remains detectable, consistently show an increased transparency toward lower energies similar to that reported previously for absorptive fluid hydrogen (Supplementary Fig. 2), suggesting that hydrogen in this regime is semiconductor-like with a band gap of the order of ~1 eV [25]. Transient reflectivity signal in this regime shows a small, spectrally independent increase (Supplementary Fig. 3) which can be explained by a small change in the refractive index of $H_2$ ($D_2$) correlated with band gap reduction[18]. In this regime peak temperature (measured radiometrically, Supplementary Fig. 4) tends to increase slowly with laser power, while the duration at which the sample remains hot (and thus emits) increases[25,26,30] (Supplementary Fig. 5); temperature increases more rapidly at higher laser power.

At temperatures exceeding 3000 K we detected a strong transient reflectivity signal from hydrogens in all samples studied (Fig. 1, Supplementary Figs. 6, 7). Reflectance of hot transformed hydrogens exceeds the background reflection substantially and is characterized by a spectrally variable magnitude (see below). At the conditions where the reflective hydrogen forms, it has a sufficiently large emissivity so its thermal radiation can be reliably collected and spectrally analyzed, enabling direct determination of the sample temperature (Supplementary Figs. 4)[25]. The reflectance spectra (Fig. 1) show a large increase to lower energy, a characteristic of metals. Within a single heating event, the reflectivity reaches a maximum when the highest temperature is reached (just after heating pulse arrival) and diminishes as the sample cools (Fig. 1b). The overall reflectance value increases with the laser heating pulse energy (and hence the maximum sample temperature) (Fig. 1c). These transient changes at high temperatures are reversible (Fig. 1a), sometimes occurring with relatively smaller changes to the background attributed to laser absorber movement; thus, they must manifest a transition in the state of hydrogens at these extreme *P-T* conditions. Raman spectra measured before and after heating events demonstrate the vibron mode of hydrogen and do not show any extra peaks that could be related to irreversible chemical transformations that would occur as a result of exposure of hydrogen to extreme *P-T* conditions.

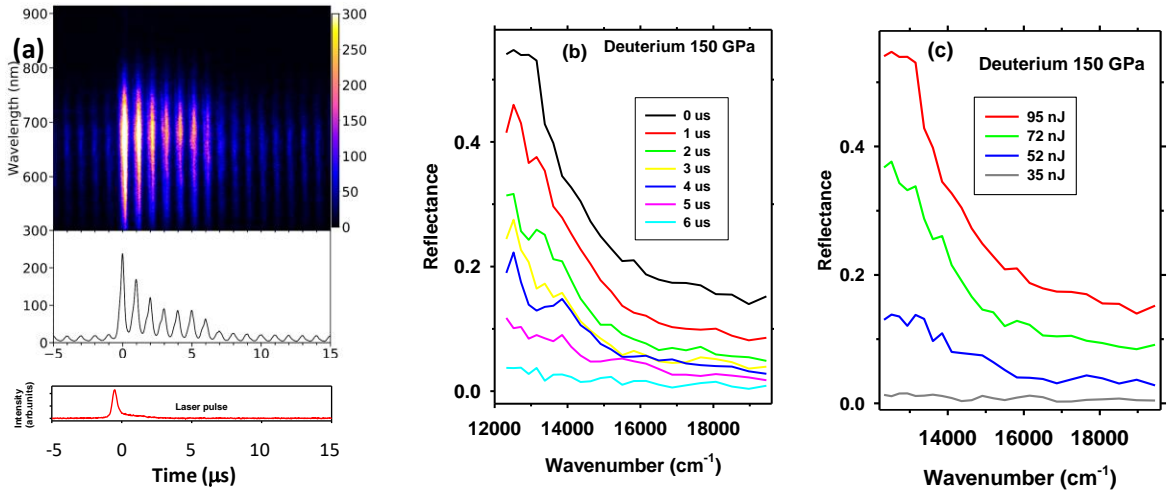

**Figure 1. Transient optical reflectivity data of deuterium at 150 GPa.** Left panel (a): spectrogram of time dependent reflectance of laser heated deuterium at 150 GPa using SC pulsed probes dispersed by a diffraction grating and recorded via a steak camera. The SC pulses are arriving with a 1 μs time interval. The laser heating pulse arrives just before the 0$^{th}$ μs SC laser pulse. Middle panel (b): the transient reflectance spectra measured at the different times after the arrival of the heating pulse, probing the sample at varying temperature; Right panel (c): maximum reflectance as a function of laser pulse energy. Temperature is 4400(600) K at peak heating (panel (b)).

Our experiments probe the electronic properties of conducting fluid hydrogen by examining its response to optical radiation in the visible spectral range. Hydrogen in molecular fluid state is insulating, so it is transparent in the spectral range of investigation and it reflects lights at the interface with another material due to a difference in the refractive indices (Fresnel reflection). The room-temperature refractive index of hydrogen rapidly increases with pressure[34] and in the range of this investigation it is close to that of diamond and alumina (which change much slower due to substantially smaller compressibility) so the "background" reflectivity on the insulating sample interface cannot exceed 4%; in fact, we detected less than 1% in all our experiments. The secondary reflections, proximity of a heat absorber, and sometimes interference effects can slightly increase this value but it still remains low compared to the response of metallic hydrogens. This background has been subtracted (Supplementary Fig. 6) from the transient reflectivity spectra. We stress that since this background reflectivity always has a relatively small contribution, the ambiguities related to the removal procedure could not affect our results substantially. The absolute reflectance values varied in different experiments, depending on the achieved temperature, the sample geometry (absorber material and diameter of the hole in it) which affects the shape of the metal-insulator interface and orientation relative to the probe beam; nonetheless, the spectral characteristics remain very similar (Supplementary Fig. 7). The presence and optical depth of any semiconducting absorptive hydrogen at lower temperatures adjacent to the metallic state can also shield the metallic state from direct probing[31]; however, metallic hydrogen can here be presented solely in very thin (*e.g.* sub μm) layers between the heat absorbers (*e.g.* boron doped diamond) and alumina coatings (**Methods**), which was likely the case of higher reflectivity measurements in deuterium (Figs. 1(b,c)).

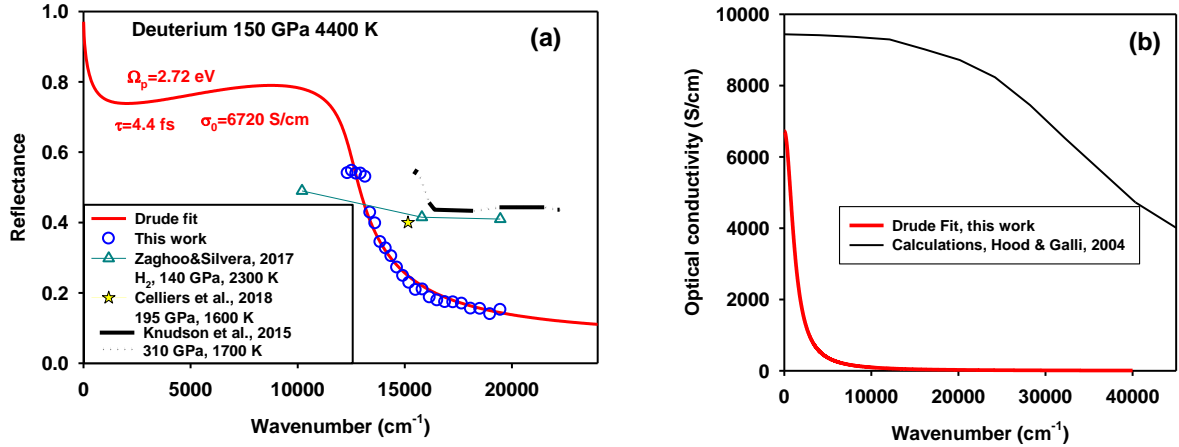

**Figure 2. Optical conductivity of metallic deuterium.** (a): Reflectance spectrum of deuterium; the experimental data (circles) are compared to the Drude fit (**Methods**), which yields DC conductivity; it compares well to the theoretical values[14,15,35-38] (near 10,000 S/cm) as well as dynamic gas gun (2,000 S/cm)[10], pulsed electrical generator (>10,000 S/cm), [13] and laser driven compression (>10,000 S/cm)[12] experiments. Shown for comparison, albeit at different thermodynamic conditions, are results of DAC experiments for hydrogen [11] (triangles) using three monochromatic laser probes, and laser driven dynamic experiments for deuterium that used passive spectroscopy [13] (solid black lines with dotted grey interpolation) and monochromatic probing (star) [12] (b): Optical conductivity from this work compared to that theoretically computed in ninefold compressed deuterium at 3000 K [15].

The reflectance measurements of hydrogen all yielded qualitatively similar spectra (Supplementary Fig. 7) with the pronounced increase in intensity toward low energy. These spectra can be fitted with a variety of different models, but we find that a Drude free electron model (**Methods**), which employs the plasma frequency $\Omega_P$ and the mean free time between the electron collisions $\tau$ as the free parameters, fits the data well, yielding $\Omega_P$ =2.72(5) eV and $\tau$=4.4(1.6) fs for deuterium, where the detected reflectance was largest (Fig. 2). In these calculations, we assumed that the refractive index of warm nonmetallic hydrogen in contact with metallic hydrogen is 3.0 at extreme *P-T* conditions following recent dynamic compression measurements[12]. Furthermore, our reflectivity data in the high frequency limit can only be accurately fitted by including a bound electron contribution to the electronic permittivity function of metallic hydrogen ($\varepsilon_b$ =3.1 for the representative case above). The uncertainty of our estimation of the DC conductivity $\sigma_{DC}=\Omega_P^2\tau$ is of the order of 30%, $\sigma_{DC}$=6700(2400) S/cm.

The reflectance spectra (Fig. 1(b,c)) at various temperatures varied either during cooling down or by changing the laser heating energy can be also fit with the Drude model. The results of time domain experiments on cooling down show a change in a slope in the Drude parameters at the critical onset temperature $T_c$ (Fig. 3), where the reflectance becomes less than approximately 10%. The most prominent change is in the DC conductivity, which is almost constant above the onset transition (although the reflectance values vary) and start dropping down fast below $T_c$ manifesting the transition. This is qualitatively similar to the recently reported behavior of deuterium under ramp compression near 200 GPa, albeit probed as a function of pressure [12]. We also find that $\tau$ decreases from the metallic state through the transition (Fig. 3). Furthermore, we find that the electronic permittivity $\varepsilon_b$ increases and although the plasma frequency increases in the metallic state, the "screened" plasma frequency $\Omega_P/\sqrt{\varepsilon_b}$ remains constant and drops in a semiconducting state (Supplementary Fig. 8). These observations suggest an electronic oscillator frequency shift from high energy toward zero as metallization progresses, which is further supported by our optical absorption data (Fig. 4), in the regime of low reflectivity. This is a common feature in insulators undergoing metallization (*e.g.* Ref. [39]) resulting from charges becoming increasingly less bound, while the scattering time also increases considerably into the metallic state, which can also be attributed to a transformation from localized to delocalized carriers.

The DC (electrical) conductivities inferred here are in reasonable agreement with the results of theoretical calculations (ca 10,000 S/cm)[14,15,17,35-37] and compares well to the dynamic experiments on metallic hydrogens[10,12]. However, our experiments suggest a more than an order of magnitude longer electronic collision time $\tau$ in the metallic state implying that the conducting electrons in hydrogens at the conditions studied are not in the MIR limit. In the absence of the spectral reflectance data, the validity of the MIR conditions was a common assumption in analyzing the dynamic compression data[12,18,19]; theoretical calculations were in a general agreement predicting a very damped Drude response[15,38] (Fig. 2(b)). Our reflectivity spectra are in partial disagreement with those reported in the dynamic experiments[13] (Fig. 2(a)), though these refer to substantially different *P-T* conditions (Fig. 5), do not cover a near IR spectral range, are obtained on a hydrogen-LiF interface, and use a passive spectroscopy technique sensitive to diffuse scattering, making a direct comparison of reflectance values possibly inappropriate; however, some evidence of a sharper rising reflection to lower energy, similar to that observed here, is noted in these data. Recent DAC experiments[11] at similar conditions to the present results report a value of $\sigma_{DC}$=11,000 S/cm, which agrees broadly with our determination, but the reflectance results differ drastically (Fig. 2(a)): a Drude fit to those data [11] yielded a larger plasma frequency ($\Omega_P$ =20.4 eV) and smaller electron collision time ($\tau$=0.13 fs) compared to our results. The distinction between our results and those of Ref. [11] are unlikely due to a difference in the probed *P-T* conditions. In fact, we find the onset of metallic conditions at higher temperature than in Ref. [11,26], in better agreement with the results of dynamic experiments[10,12,13] (Fig. 5) with regard to the *P-T* conditions of metallization. The differences in inferred metallization conditions and spectral response may be due to the larger background signal in Refs. [11,26], from a tungsten layer in the probed sample region, the optical properties of which at extreme *P-T* conditions are unknown.

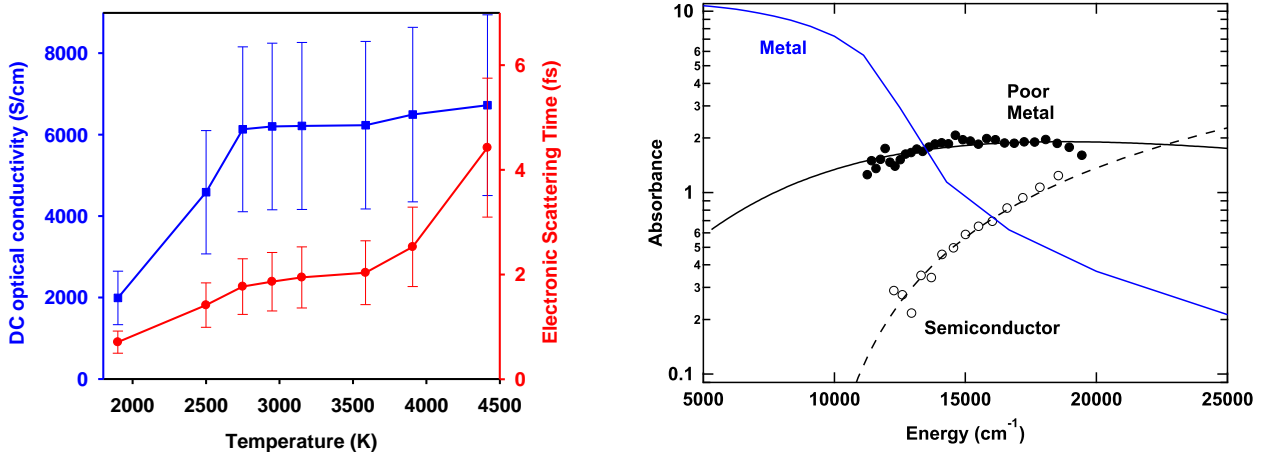

**Figure 3. Optical conductivity and electronic scattering time through metallization of deuterium at 150 GPa.** The results are obtained in the same experiment on cooling down from the top temperature. Temperatures below 3000 K are determined via linear extrapolation with time.

**Figure 4. Absorption spectra through metallization at 140-150 GPa.** Spectrum of metallic deuterium at 150 GPa, 3000 K determined from Drude fit to reflectivity and having $\sigma_{DC}$=6700 S/cm, $\tau$=4.4 fs is the solid blue line (after Fig. 2). Measured spectrum of a transitional, poor-metal deuterium state at 150 GPa, T<2800 K, given by the red points (Fig. S2(c)), is best fit by a Smith-Drude model (solid black line), here with $\sigma_{DC}$=35 S/cm, $\tau$=0.3 fs, *C*=-0.95. Prior data on semiconducting hydrogen[25] at 141 GPa, 2400 K (open symbols) are best fit by a Tauc model (dashed black line) with gap energy of ~1 eV, and corresponding to $\sigma_{DC}$≈15 S/cm, $\tau$≈0.03 fs. A 1 µm thick layer is assumed in calculations.

The sharp reflection rise in the visible spectral range documented here is remarkable. We assign it to the presence of a plasma edge common for many metals, for example gold and silver. Such electronic excitations with the frequencies near the plasma edge are not unusual for simple metals; these would represent the electronic transitions to excited bound states, which could correspond to weakly bound dimers of hydrogens. In this regard, we have attempted to reproduce our reflectivity spectrum by using a two-oscillator model. In this model one oscillator is at zero frequency (Drude term) and another is at the frequency beyond the observation limit (Supplementary Fig. 9). To match the observations, the free electron plasma frequency is $\Omega_P$ =6.7 eV, and the second oscillator should be at nearly 25,000 cm$^{-1}$. However, the DC conductivity in this model must be near $\sigma_{DC}$=61,000 S/cm, an order of magnitude larger than for the Drude model, which is inconsistent with the dynamic electrical conductivity experiments [10]. We thus prefer a simple description of our data with a free electron model.

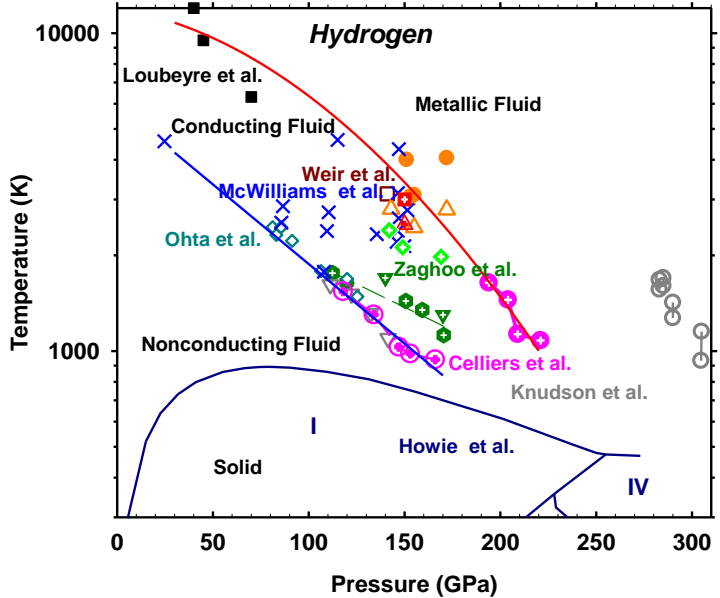

**Figure 5. Phase diagram of hydrogen at extreme *P-T* conditions.** Filled orange circles indicate conditions of the metallic state detected via optical reflectance in this study for hydrogen (filled crossed red square for deuterium). Open orange (crossed red for deuterium) upward triangles correspond to P-T conditions where hydrogen reflectivity was lower than a few percent and our Drude analysis shows a sharp decline in the DC conductivity (Fig. 3). The uncertainty of temperature measurements is about 15% (1σ). Blue crosses are the conditions of absorbing hydrogen directly measured using a similar DAC technique as in this work [25]. Open and filled pink crossed circles are the results of laser shock experiments at NIF corresponding to reaching the absorptive and reflecting deuterium states, respectively [12]. Open brown square is the result of reverberating shock experiments detecting metallic hydrogen by electrical conductivity measurements [10,13]. Filled black squares are IMT measured in single-shock experiments in precompressed samples [19]. Open gray triangles and circles are the results obtained in dynamic compression experiments on the Z-machine that correspond to absorptive and reflecting deuterium [13]. The results of DAC experiments reported as an abrupt insulator-metal transition in hydrogen and deuterium are shown by various green symbols [11,26,29]. Open dark cyan diamonds are the results of DAC experiments in hydrogen where a change in the temperature vs laser power slope indicate phase transformation to a metal [27,28]. Solid red and blue line through the data are the suggested phase boundaries for semiconducting and metallic hydrogens. The melting curve and solid state boundaries are from Ref. [40].

The results presented here confirm and reestablish the existence of two transformation boundaries corresponding to the formation of absorptive and reflecting hydrogens (Fig. 5). The one at lower *P-T* conditions has been established in dynamic[12,13] and DAC experiments [25,27,28,30]. It has been suggested that this boundary is related to a band gap closure [12,13,25], rather than the plasma transition. However, the absorption edge is broad (~1 eV) [25], while the transition is rather abrupt (a few hundreds of degrees); such large temperature driven bandgap changes are normally uncommon. Moreover, the band closure is usually treated as a pressure (density) driven transformation, while both the previous absorption and the present reflection results indicate a strongly temperature driven transition (see also Refs. [25,28]). This suggests that the observed phenomena are related to the molecular instability and the observed boundary corresponds to a temperature driven partial molecular dissociation in the regime of proximity of the molecular binding energy to the zero point energy, which can occur in the pressure range near 150 GPa [3,41]. In this interpretation with increasing temperature molecules first begin to dissociate and recombine frequently, producing a state with a measurable electrical and optical conductivity [10,12,25] but which is not metallic, with mainly localized carriers. To reach the metallic state one needs to dissociate a critical fraction of molecules (*e.g.* 40% [15]) and enable nonlocal carrier transport, which occurs at higher *P-T* conditions.

The metallization line consistent with our results is established mainly based on the results of dynamic experiments with calculated temperatures[10,12,13,19]. Our new data using direct temperature measurements show a reasonably good agreement, despite difficulties in temperature metrology on the insulating and weakly conducting states with low emissivity light absorbers, while the DAC results reported by the Harvard group (filled green crossed symbols) suggest a transition at ~1000 K lower temperatures (Fig. 5). Our results do not suggest any major isotope effect (cf. Ref. 29), which is consistent with previous shock wave results [10,19]. The lines of metallization and conductance become closer in T at higher P (Supplementary Figure 10) as expected on approaching a critical point, however the data suggest that they both would cross the melting line first. The pressure range of 170 - 250 GPa at temperatures just above the melting line can be expected to be anomalous. This *P-T* space has been probed in two recent high-temperature Raman experiments [40,42]. It is interesting that Zha et al. [42] detected an anomaly in the pressure dependence of the liquid hydrogen vibron band at 140 - 230 GPa, which can be related to the presence of conducting mixed molecular-atomic fluid hydrogen. However, they find that fluid hydrogen remains molecular at 300 GPa, which calls for more improved *P-T* metrology in dynamic laser and resistively driven static experiments (Supplementary Figure 10).

In conclusion, our investigation of fluid hydrogens in the regime of molecular dissociation and metallization showed the complexity of the phenomena suggesting a two-stage transition with a semiconducting intermediate state preceding that of a free-electron metal. The reflectance spectra of the metallic hydrogens show the presence of a plasma edge, which allow constraining the electronic transport conductivity parameters. We find an electronic relaxation time that is much larger than previously thought, suggesting unusual transport in hydrogens at extreme *P-T* conditions. Within the precision of our measurements, we find the onset of metallic conditions occurs at comparable conditions in fluid hydrogen and deuterium, consistent with theoretical expectations and the results of dynamic compression experiments.

**Methods**

**Sample preparation.** We investigated three $H_2$ and one $D_2$ samples gas loaded in the DAC with beveled diamonds of 40-70 μm diameter central culets and compressed to pressures of 120-172 GPa. Alumina coatings of a submicrometer thicknesses have been sputtered on the diamond anvil prior the gas loadings. Thin Ir or boron-doped diamond (the latter of approximately 1 μm thick) plates of 20-40 μm linear dimensions with or without small (<10 μm in diameter) cylindrical holes were positioned in the high-pressure cavity to serve as the laser radiation absorbers. Pressure was determined at room temperature using the spectral position of the Raman vibron peaks (e.g. Ref. [40]).

**Dynamic laser heating and optical probes in the diamond anvil cell.** Our time-resolved single pulse laser heating diamond anvil cell experiments combine measurements of optical emission, transmission and reflectance spectroscopy in the visible spectral range (480-750 nm) using a streak-camera, as has been described in our previous publications [25,32] (Supplementary Fig. 1). The laser pulses of 4–10 μs duration are sufficiently long to transfer heat to the hydrogen sample in the hole of the heat absorbers creating a localized heated state of several μm in linear dimensions and a few μs long as determined in our FE calculations [25]. The optical spectroscopic probes, aligned to the heated spot, were used in a confocal geometry suppressing spurious probe signals. Transient transmittance and reflectivity were obtained using a pulsed broadband supercontinuum (SC, 1 MHz, 1 ns, 480–720 nm) probe having focal spots of approximately 6 μm in diameter (Supplementary Fig. 1) that is spatially filtered with a confocal aperture of some 50% larger in diameter.

Time resolved (with the resolution down to 0.5 μs) sample temperature was obtained from fitting thermal radiation spectra emitted by the coupler and hot sample to a Planck function (Supplementary Fig. 4). These were normally determined in a separate experiment with identical heating without probing, and due to weak thermal radiation were in some cases integrated over a number of laser heating events (5-20) to improve signal-to-noise. The measured temperature should be treated cautiously as the thermal radiation measured represent a sum of contributions from the coupler and the sample. Commonly, the heat absorber has higher temperature than the sample and in the case of Ir coupler emits more because of difference in emissivity, while boron doped diamond emits very little. However, the sample emissivity changes substantially once it becomes absorptive, suggesting that the measured thermal emission in this regime characterizes the sample temperature. Additionally, FE calculations [25,32,43] have been used to model the temperature distribution in the high-pressure cavity.

At each nominal pressure, temperature was increased stepwise with an increase of the heating laser power. Pressure was measured before and after the heating cycles using Raman spectra of the hydrogen vibron and additionally the stressed diamond edge. Pressure was found to remain essentially constant (within <3 GPa) between heating cycles. No correction for the thermal pressure has been used.

**Transient optical data reduction.** To determine the transient reflectance values of conducting hydrogen at extreme *P-T* conditions we measured the reflectance of the outside diamond-air interface as a natural reflectivity standard. The background Fresnel reflection from the diamond-alumina and alumina-hydrogen interfaces were subtracted (Supplementary Figs. 3 and 6). No correction has been made for the attenuation made by an absorptive conducting deuterium (cf. Ref. [31]) due to the optimal experimental condition achieved here where the hot metallic sample is confined within a thin layer between absorber and alumina coating. The reflectivity spectra of hydrogen, which were not positioned optimally, have been used only for the qualitative analysis. While the temperature in the pressure chamber changes continuously reaching almost room temperature near the diamond anvil tip, the coating improves the sample insulation and lessens the need for absorptive semiconducting sample at lower temperature within the sample cavity, which screen the reflection from the metallic state.

**Drude model.** Reflectivity spectra are well-fitted by a Drude model having conductivity of the form $\sigma=\sigma_0(1-i\omega\tau)^{-1}$ where $\omega$ is the angular frequency. The DC conductivity is $\sigma_0=\Omega_p^2\tau$, in which $\Omega_p$ is the plasma frequency, $\tau$ the scattering time, and $\varepsilon_0$ the permittivity of free space. The dielectric constant is $\varepsilon^* = \varepsilon_b + i\sigma/\omega$ where $\varepsilon_b$ is the bound electron contribution to the dielectric constant. A range of $\varepsilon_b$ from 1 to $n_H^2$ ($n_H$ =3) are examined when assessing uncertainty; we generally found $\varepsilon_b$=3.1 provided a better fit to the data. The corresponding index of refraction is $n^* = \sqrt{\varepsilon^*}$. The absorption coefficient is determined as $2\omega*\mathrm{Im}(n^*)/c$. The reflectivity of the interface between the metallic and semiconducting hydrogen is modelled as $R = |(n^*-n_H)/(n^*+n_H)|^2$, where the refractive index of the semiconducting state (n=3.0) is taken from Ref. [12], where it was measured in very similar thermodynamic conditions in laser driven ramp compression experiments. The Smith-Drude model has been used following the formulations in Refs. [12,25], i.e. $\sigma=\sigma_0(1-i\omega\tau)^{-1}[1+C(1-i\omega\tau)^{-1}](1+C)^{-1}$.

**Data availability.** All relevant numerical data are available from the authors upon request.

**Bibliography.**

# Supplemental information

# To

# Insulator-metal transition in liquid hydrogen and deuterium

Shuqing Jiang[1,2], Nicholas Holtgrewe[2,3],†, Zachary M. Geballe[2], Sergey S. Lobanov[2,4],‡,

Mohammad F. Mahmood[3], Alexander F. Goncharov[1,2]∗

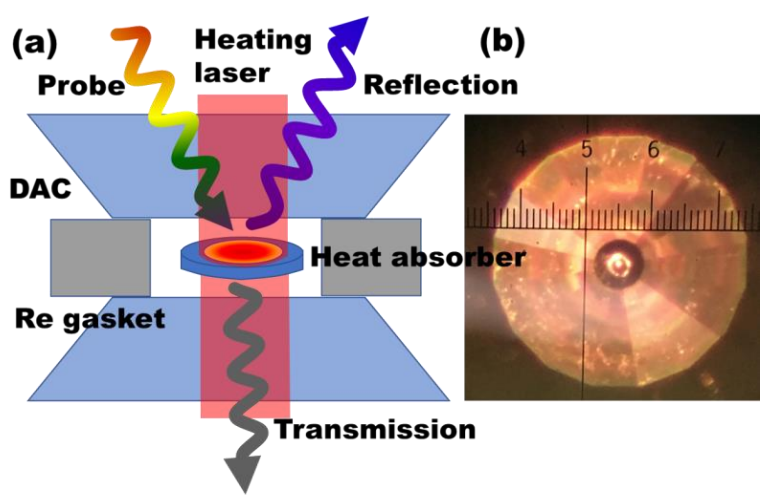

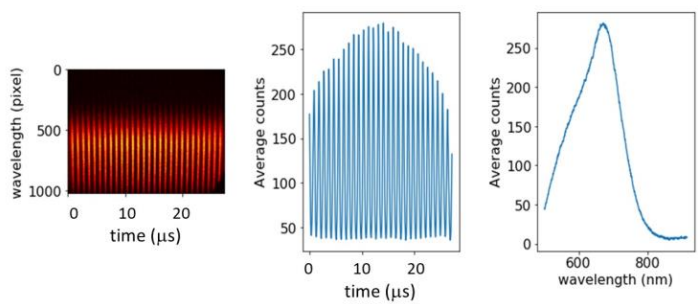

**Supplementary Figure 1. Experimental geometry of a diamond anvil cell experiment along with a microphotograph (top) and characteristics of a spectral probe (supercontinuum) laser measured via a streak camera (bottom).**

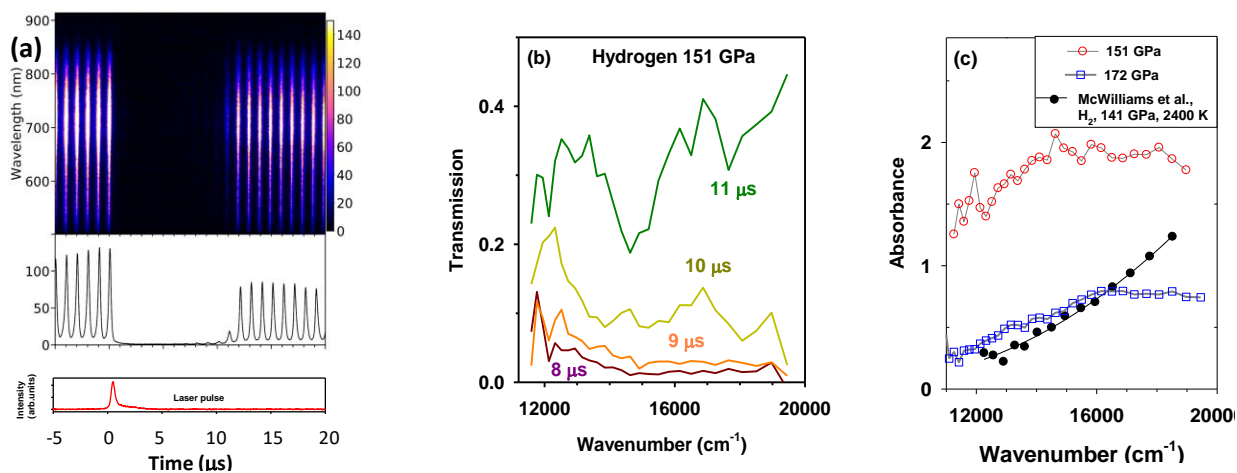

**Supplementary Figure 2. Transient optical transmission data of hydrogen at 151 and 172 GPa.** Left panel (a): transmission spectrogram at 151 GPa, which shows the time dependence of the pulsed supercontinuum (SC) signal transmitted through the sample during pulsed laser heating (bottom panel). It is dispersed by a diffraction grating and recorded via the streak camera. The SC pulses are arriving with a 1 μs time interval. The laser heating pulse shown in the bottom arrives after the SC laser pulse at the $0^{th}$ μs. The temperature was too low (<3000 K) to be detected in this single event experiment. The middle panel (b) shows the transmission spectra at different times and the right panel (c) depicts the optical absorption spectra at 151 and 172 GPa (on cooling) in comparison to that reported in Ref. [25].

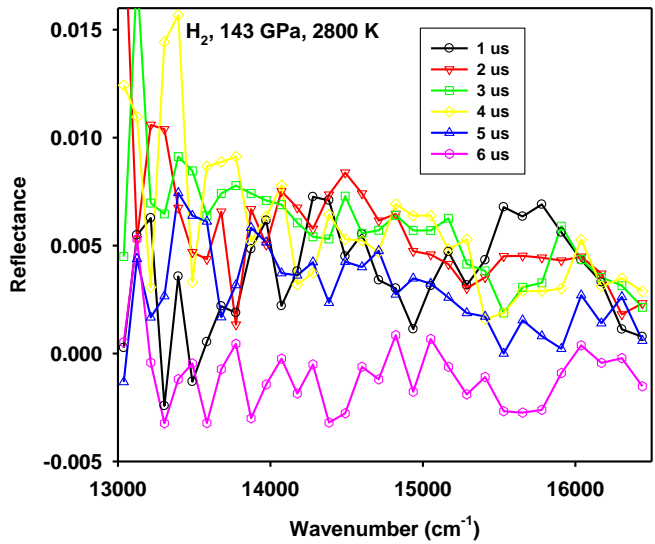

**Supplementary Figure 3. Reflectivity spectra at 143 GPa.** The heating pulse that arrives near the $0^{th}$ microsecond causes a small uniform increase of reflectivity. Each spectrum is measured using a single SC pulse, which probes the sample every 1 µs. The spectra are normalized to the reflectance of the diamond-air interface (the anvil table) and the background signal was subtracted.

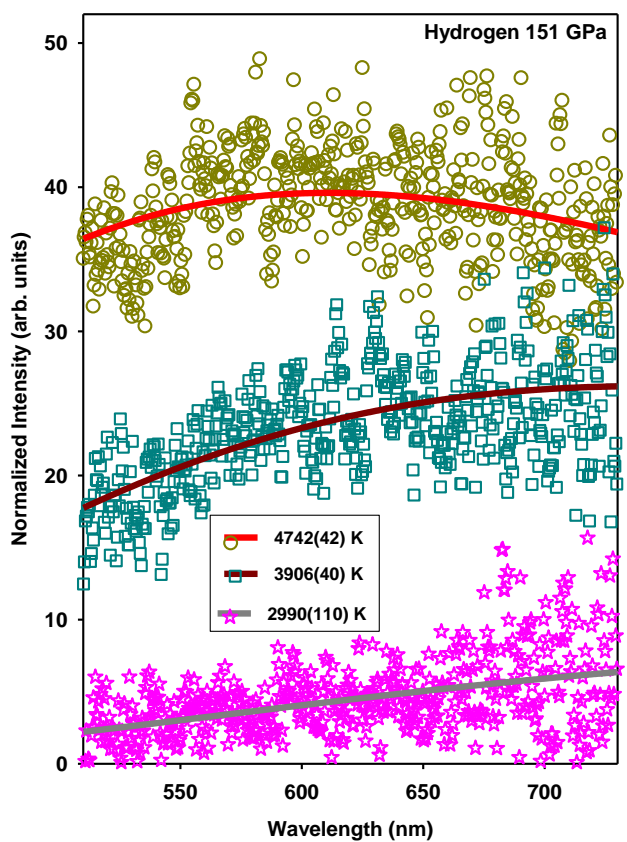

**Supplementary Figure 4. Spectroradiometric transient temperature measurements for hydrogen at 151 GPa.** The data for various representative times are obtained in the course of Heat 5 (Suppl. Fig. 5).

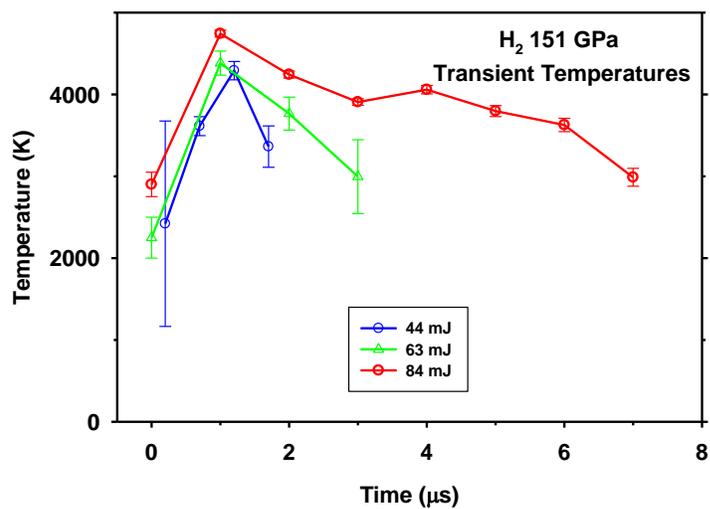

**Supplementary Fig. 5.** Transient temperature measured with spectroradiometry at 151 GPa with hydrogen as a sample. Reflectivity measurements showed a very small increase for Heats 3 (44 mJ) and 4 (63 mJ), while a stronger reflectivity with a characteristic increase toward the lower energy side was detected in the course of Heat 5 (84 mJ).

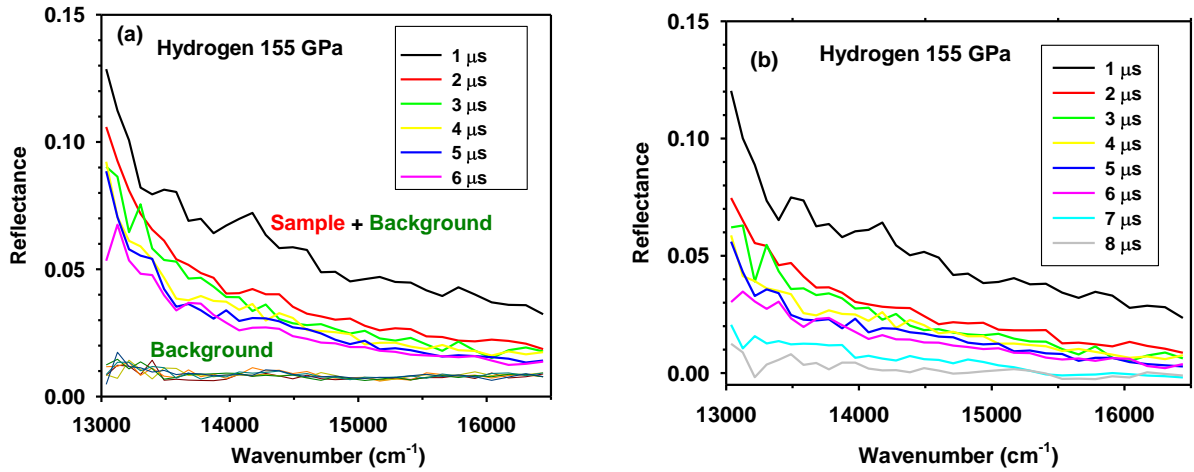

**Supplementary Figure 6. Reflectivity spectra of hydrogen at 155 GPa.** The heating pulse arrives near $0^{th}$ microsecond. Left panel (a): raw data of the reflectance measured using SC pulses before the heating (background); Right (b): the background is subtracted.

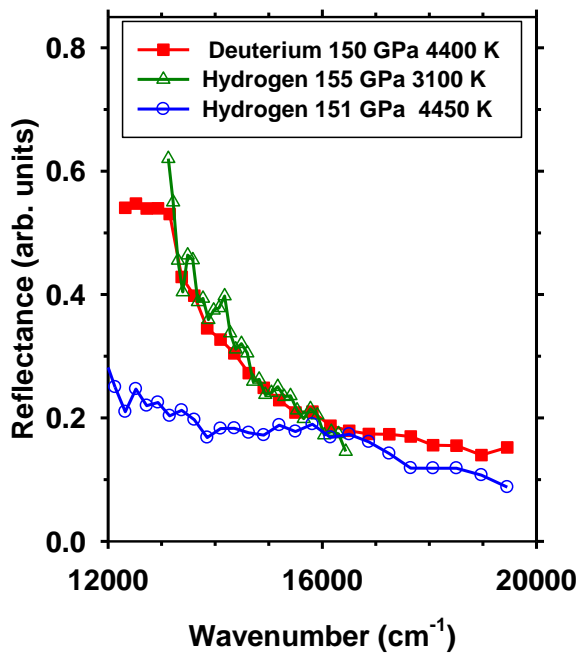

**Supplementary Figure 7. Reflectivity spectra of hydrogens in different experiments.** The spectra are scaled to roughly match the reflectivity values at about 15000-20000 cm$^{-1}$.

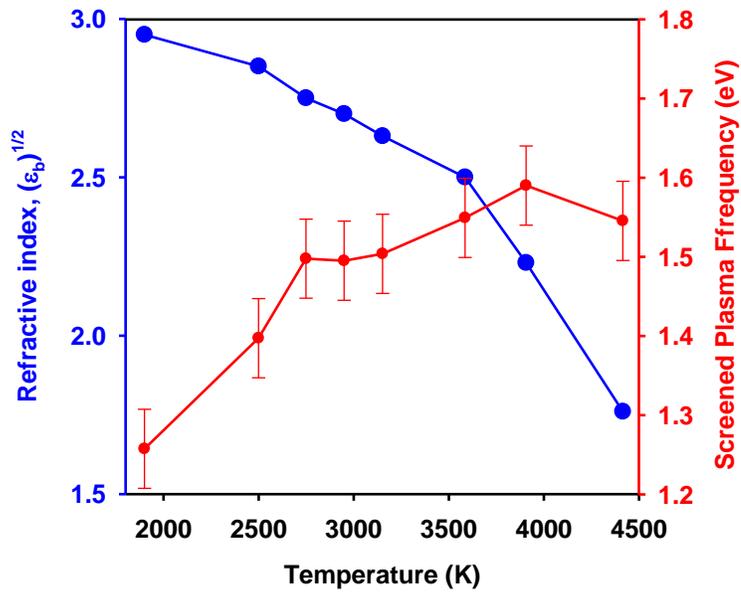

**Supplementary Figure 8. Parameters of the Drude model that have been used to fit the optical reflectivity data of Fig. 3.** The results are obtained in the same experiment on cooling down from the top temperature. Temperatures below 3000 K are determined via a linear extrapolations as a function of time.

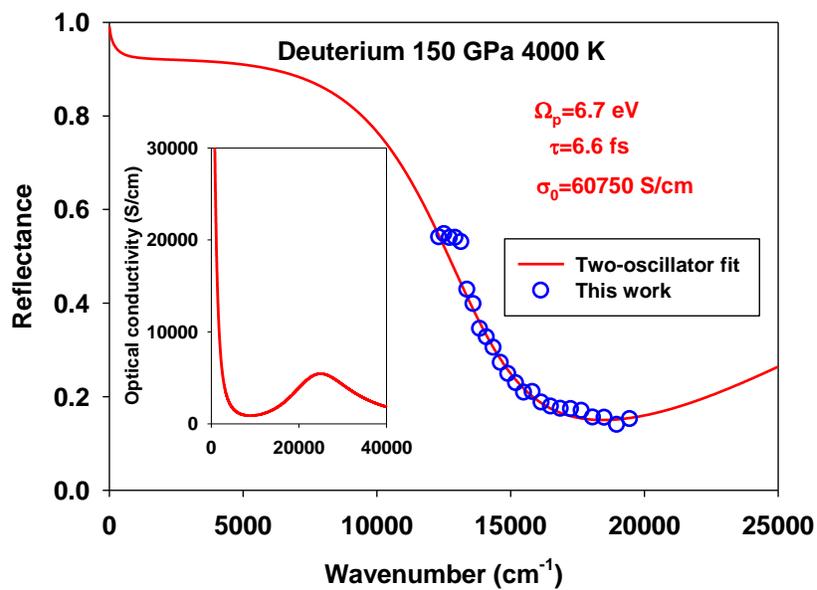

**Supplementary Figure 9. Reflectivity spectrum of metallic deuterium fit using a model of two Lorentz oscillators.**

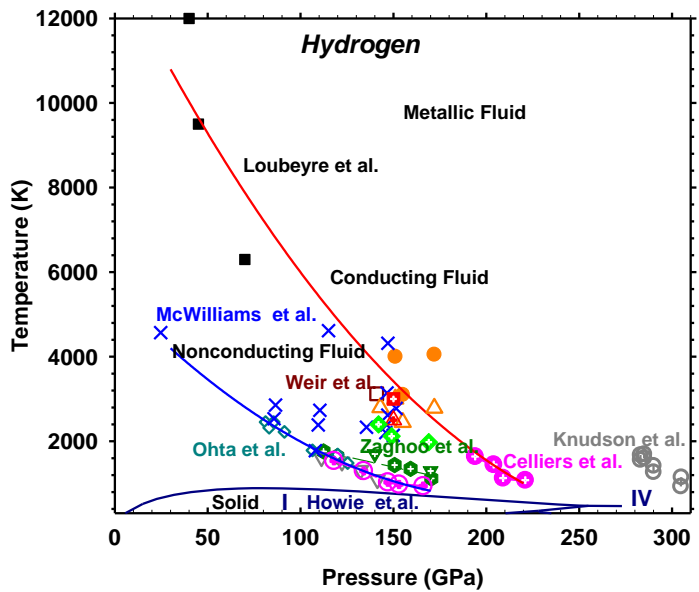

**Supplementary Figure 10. Phase diagram of hydrogen at extreme *P-T* conditions (in linear coordinates). Phase diagram of hydrogen at extreme *P-T* conditions.** Filled orange circles indicate conditions of the metallic state detected via optical reflectance in this study for hydrogen (filled crossed red square for deuterium). Open orange (crossed red for deuterium) upward triangles correspond to P-T conditions where hydrogen reflectivity was lower than a few percent and our Drude analysis shows a sharp decline in the DC conductivity (Fig. 3). The uncertainty of temperature measurements is about 15% (1σ). Blue crosses are the conditions of absorbing hydrogen directly measured using a similar DAC technique as in this work [25]. Open and filled pink crossed circles are the results of laser shock experiments at NIF corresponding to reaching the absorptive and reflecting deuterium states, respectively [12]. Open brown square is the result of reverberating shock experiments detecting metallic hydrogen by electrical conductivity measurements [10,13]. Filled black squares are IMT measured in single-shock experiments in precompressed samples [19]. Open gray triangles and circles are the results obtained in dynamic compression experiments on the Z-machine that correspond to absorptive and reflecting deuterium [13]. The results of DAC experiments reported as an abrupt insulator-metal transition in hydrogen and deuterium are shown by various green symbols [11,26,29]. Open dark cyan diamonds are the results of DAC experiments in hydrogen where a change in the temperature vs laser power slope indicate phase transformation to a metal [27,28]. Solid red and blue line through the data are the suggested phase boundaries for semiconducting and metallic hydrogens. The melting curve and solid state boundaries are from Ref. [40].